\documentclass[prb,preprint%
,showpacs]{revtex4}
      \usepackage{graphicx}

\usepackage{dcolumn}
\usepackage{bm}
\begin{document}
\title{Inhomogeneous gas model for electron mobility in  high density neon gas}

\author{A.F.Borghesani}\email{borghesani@padova.infm.it}
\affiliation{Istituto Nazionale per la Fisica della Materia \\
 Department of Physics, University of Padua\\
 Via F. Marzolo 8, I--35131 Padua, Italy}
 
\author{T.F.O'Malley}\email{tfomalley@earthlink.net}
\affiliation{24724 Valley St., 304, Santa Clarita, CA 91321, U.S.A.}%

\begin{abstract}
   Experimental studies of electron mobilities in Neon as a function 
   of the 
   gas density have persistently shown 
   mobilities up to an order of  magnitude  
   smaller  than expected and predicted.   
   A previously ignored  mechanism 
   (gas in--homogeneity which is negligible in the thermal mobilities for He and 
   other gases) is found  to reproduce the observed Neon mobilities 
   accurately and simply at five temperatures with just one variable parameter. 
   Recognizing that a gas is not a homogeneous medium, 
   a  variation in local density combined with the quantum multi--scattering 
   theory, shifts the energy and cross section -- 
   which in turn changes the collision probability and finally the mobilities.  
   A lower density where a momentum transfer interaction occurs moves 
   the mobility strongly in the same direction as the anomalous experiments.  
   By going backwards from the observed mobilities, the collision 
   frequency at each temperature and density is made to reproduce the 
   experimental data by looking for the local (as opposed to average) 
   density at which the (rare) momentum transfer interactions occur. 
   These density deviations give a picture of the size and behavior 
   of the wave packets for electron motion which looks very much like 
   the often discussed wave function collapse.
\end{abstract}
\maketitle

 \section{Introduction}\label{sec:intro} 
The study of the response of excess electrons in gases under the action of
an externally applied electric field has attracted the attention of many
researchers since very early times.  On one hand, at quite low density,
electron mobility measurements have been carried out to determine the
electron-atom momentum-transfer scattering cross section in the framework
of the traditional approach of the Boltzmann equation.

	On the other hand, the transport properties of electrons in dense 
	gases~\cite{christo84} and
	liquids~\cite {holsch89} have also been and still are subject of extensive studies because
	of their potentiality in technologically relevant applications, such as
	high-energy particle detectors, as well as for the rich wealth of
	information they can provide on the basic physical mechanism of
	electron-atom interaction in disordered and condensed media.

     The mobility of extra electrons can be used as a probe to investigate the
     nature, energetics and dynamics of electron states in a disordered medium
     and their evolution as a function of the gas density.  In particular, the
     research is aimed at investigating the transition from classical single
     scattering in the completely dilute phase to multiple scattering and weak
     localization at low and moderate densities and finally leading to the
     formation of fully localized\cite{hern91} or extended\cite{borgpatrasso} states in the condensed 
     phase.

	Dense gases are the simplest realization of dense disordered systems and
	the problem of the electron motion in a dense environment of randomly
	located scatterers is a model problem for several phenomena in various
	areas of physics, such as the physics of doped semiconductors, non polar
	liquids, electrolytes in solution, non ideal plasmas.

	The classical theory of scattering in the binary collisions approximation
	predicts that the electron mobility $\mu$ depends on the nature of the
	electron-atom interaction potential through the momentum transfer
	scattering cross section, and also on the gas temperature, on the applied
	electric field $E,$ and that it is inversely proportional to the gas 
	density $N,$ 
	so that the so called density--normalized mobility $\mu_{0} N$ at thermal energy
	(i.e., at zero electric field) does not depend on the 
	density\cite{hux}.

	However, experiments have revealed anomalous density effects on the
	electron zero-field mobility.  It is now well established that in certain
	gases $\mu_{0} N$ decreases with increasing $N,$ thus showing a negative density
	effect, while in other gases the anomalous effect is positive, i.e., 
	$\mu_{0} N$
	increases with $N.$

    The amount of the density effect is a function of several factors,
    including gas density and temperature.  Among the noble gases, $\mu_{0} N$ decreases
    by a factor $\approx  10 $ in He\cite{bartelsHe} at $T=77.4$ K in the range up to 
    $N\approx  60 \times  10^{26}$ m$^{-3},$
    whereas it increases by a factor $\approx 30$ in Ar at $T=152$ K in 
    an extended density range\cite{borg2001}
    going up to $N\approx 100\times 10^{26} $ m$^{-3},$ or a little 
    bit less at higher temperatures\cite{bartelsar,borglamp}.

    The sign of the density effect on the electron mobility has been found
    experimentally to depend on the sign of the electron-atom scattering length
    $ A,$ which distinguishes whether an atom attract or repels low energy
    electrons.  It is observed that the effect has a sign in the opposite
    direction to $A.$ The effect is positive for attractive gases such as Ar with
    $A < 0$ and negative for He and Ne with $A > 0. $ The reasons for this are not
    immediately obvious and depend on the way that the quantum multiple
    scattering theory affects the mobility.
  						
   The multiple scattering theory (MS) of waves\cite{lax,foldy,keller64} treats the environment in
   which a propagating electron wave is immersed as a
   Fermi's\cite{fermi34} infinite
   sea of atoms which acts as a potential well or barrier depending on whether
   the atoms attract or repel low energy electrons. At finite 
   densities MS is a source of potential scattering, as distinct from 
   the (relatively rare) momentum transfer interactions which change 
   the electron's total asymptotic energy.

   The real part $\Delta$ of its effect raises or lowers the electron's kinetic energy level accordingly. 
   Its imaginary part $\Gamma$ acts rather as a quantum energy uncertainty or width
   which is associated with possible barrier tunneling.

     The density effect in attractive gases is a straightforward application of
     multiple scattering theory.  Starting with the classical equation for the
     mobility given by Huxley and Crompton~\cite{hux} and assuming the gas to be a
     homogeneous medium, one simply adds the shift to the electron energy.

   When MS shifts the electron energy, this change is passed directly from
   energy to cross section to collision frequency -- to electron 
   energy distribution function -- and finally to mobility. 
   For Ar these changes predicted the observed density effects always in the
   direction found by experiment and with at least fair accuracy.  However the
   effect for repulsive gases is not so simple and direct.

   In the repulsive gases at the highest densities and low temperature, the
   electron-atom repulsion may be so effective as to give origin to the
   formation of electron states localized inside fluid dilations named
   bubbles, as it is well known in liquid He\cite{meyer58} and 
   Ne\cite{bruschi72}, dense He gas\cite{LS} and dense Ne 
   gas\cite{borg90}.  No such an effect has
   been observed in gases with a positive density effect.
 \begin{figure}[htbp]
    \centering
    \includegraphics[scale=0.5]{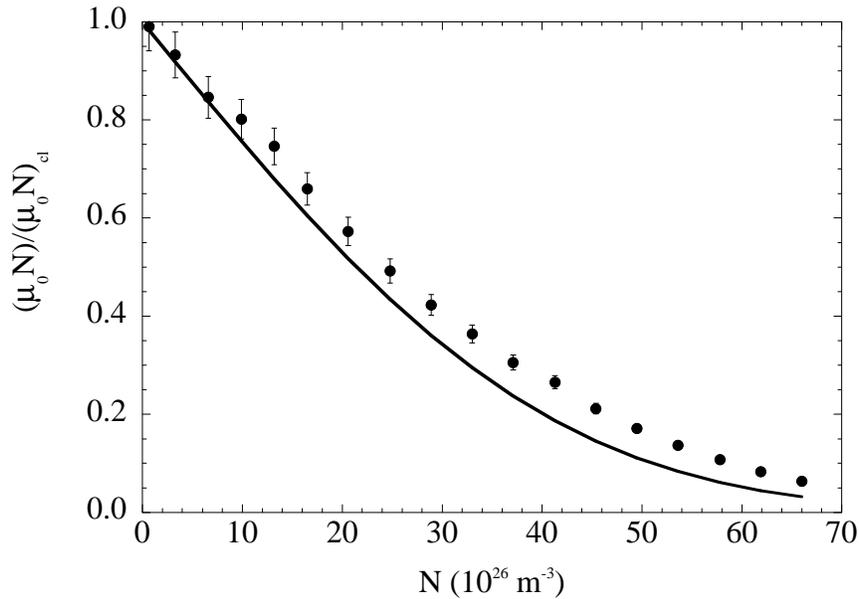}
    \caption{\small Experimental zero--field 
    density--normalized mobility ratio $\mu_{0}N/(\mu_{0}N)_{cl}$ 
    data in He gas at $T=77.4 $ K 
    \cite{bartelsHe} (points) and the prediction of MS theory 
    for a homogeneous gas\cite{om80} (solid line).}
    \label{fig:fig1}
\end{figure}
   The equat
   For repulsive gases like He and Ne at low and moderate densities (the
   subject of present interest) it is now commonly accepted that multiple
   scattering theory has the effect of defining a mobility 
   edge\cite{mott67} 
   such that electron states with large wavelength compared to their mean free
   path are effectively localized and do not propagate.
 
   The mobility edge, as found by OÕMalley in 1980\cite{om80} and from a different point
   of view in 1992\cite{om92}, is defined in Sec. \ref{oldtheories} below with the corresponding
   equations for electron mobility.  Alternative theories involving a mobility
   edge have also been advanced by other 
   researchers~\cite{poli,vdc,atra77}.
ions shown in Sec. \ref{oldtheories} below predicted fairly closely the
   negative density effect 
   in He, H$_{2},$ and CO$_{2}$ and were expected to be valid for
   any repulsive gas.
 Figure 1 shows a typical case of the predicted and observed density effect
 in He at 77K.
\begin{figure}[htbp]
    \centering
    \includegraphics[scale=0.5]{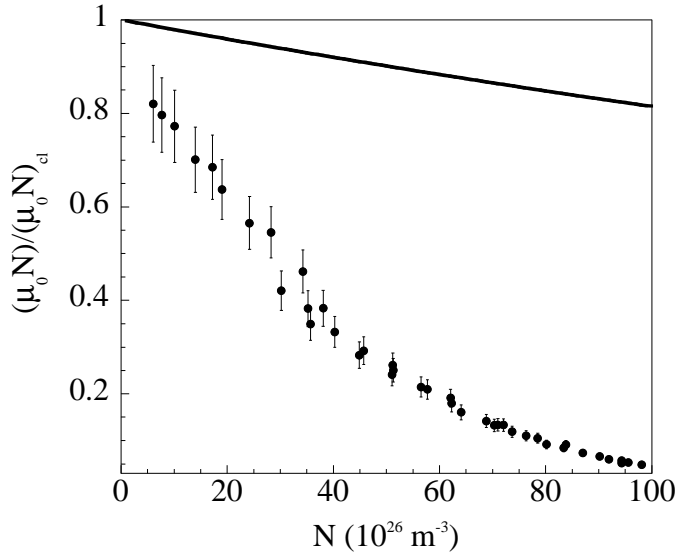}
    \caption{\small Experimental zero--field 
    density--normalized mobility ratios $(\mu_{0}N)/(\mu_{0}N)_{cl}$ data 
    in Ne gas at 
   $T\approx 46.5, $ and $T=47.9$ K~\cite{borg88,borg90} (points). 
   Solid line: prediction of MS theory\cite{om80} for a homogeneous 
   gas.}
    \label{fig:fig2}
\end{figure}   \mbox{}
   However, when the Neon data appeared\cite{borg90,borg88}, %
   the relative
   values of its density--normalized mobilities for T=46.5 and 47.9K were
   reduced by a full order of magnitude beyond what was predicted, and this
   disagreement persisted to room temperature.  The low temperature situation
   is shown in Figure 2 where the observed mobilities relative to the
   classical values $(N=0)$ are shown.

  This strong disagreement shows that the theory was seriously insufficient
  for the special case of Neon and presents a serious challenge to theory if
  an unified picture is to be maintained for the physical process of electron
  scattering off noble gas atoms in a dense environment.

  In Section \ref{neoninhom} we reexamine the generally overlooked assumption that a gas
  may be always treated as a homogeneous medium when electrons transfer
  momentum.  What we have found is that the Neon experiments themselves tell
  us (through the collision probability $\nu$ ) that the assumption is not valid
  for Neon and that $\nu$ actually favors momentum transfers occurring at
  densities less than the average.

\section{Existing theory for the thermal mobility 
in a gas as a homogeneous medium}\label{oldtheories}

We start with an existing theory\cite{om80,om92}
for drift and diffusion in a gas considered as homogeneous. Its basic component is the
Boltzmann theory of Huxley and Crompton\cite{hux} which treats the electrons as
freely propagating classical particles with the energy balance determined
by their momentum--transfer collisions, but with the collision frequency at
the point of momentum transfer determined from the quantum mechanical
cross section.

 The effect of elevated density is first included through quantum multiple
 scattering theory~\cite{lax,foldy,keller64},
 a generalization of the Fermi energy shift\cite{fermi34}.  
 The theory predicts a complex shift in a
 free electron's (kinetic) energy $\epsilon$ whose value is

\begin{equation}
    \Delta\epsilon = \Delta +i\Gamma
    \label{eq:eq1}
    \end{equation}
    where
\begin{equation}
    \Delta = 4\pi \mathtt{Re}\left[f\left(0\right)\right] = - 4\pi N A
    \label{eq:eq2}
    \end{equation}    
and 
\begin{equation}
\Gamma = \hbar N\sigma v    \label{eq:eq3}
    \end{equation}   
Quantities are generally in atomic units, (a.u.) $(m=e=h=1)$ with 
energy in Ry units (energy = $p^{2} = v^{2}$). $f(0)$ 
is the forward electron-atom scattering amplitude, $A$ is 
the scattering length, $\sigma$ is its
momentum transfer cross-section, and $v$ is the electron's classical velocity 
$(v=p/m).$

    For the present case of Neon, a repulsive gas (negative scattering length
    $A$) as is He, the problems created by shifting the lowest electron energy in
    the negative direction was solved in terms of a mobility edge 
    $E_{c}$~\cite{mott67}, 
    and resulted in the following equation for the electron's 
    density--normalized mobility $\mu N$

\begin{equation}
    \mu N = -const \int\limits_{0}^{\infty}
    \left[ {p\over \nu\left( p\right)}\right] 
    {\mathrm{d}g\over\mathrm{d}\epsilon}
    p^{2}\,\mathrm{d}p
    \label{eq:eq4}
\end{equation}
where 
\begin{equation}
    g(\epsilon) = const \int\limits_{0}^{\epsilon} 
   \left[ k_{\mathrm{B}}T+ {C\over \nu^{2} \left( p\right)}\right]^{-1}
   \mathrm{d}\epsilon_{0}\label{eq:eq4bis}
   \end{equation}
   where $C= e^{2}(M/m)E^{2}/3$. Eq. \ref{eq:eq4bis} defines the distribution 
   function\cite{lek67}
   $g(\epsilon)$ for the electron's energy of propagation $\epsilon  = 
   p^{2}$ in Rydberg units, where $p$ is the electronÕs momentum.

   The initial $p$ in the integrand of Eq. \ref{eq:eq4} is the velocity (in a.u.), whose energy
   integral determines the drift velocity $v_{D}= (\mu N)\times( E/N).$ 
   It effectively cancels
   the $p$ factor in the collision frequency $\nu$ (Eq. \ref{eq:eq8} below) so that the
   mobility in a repulsive gas depends effectively on $1/\sigma$ rather 
   than on $1/(\sigma
   v).$
   
   In Eq. \ref{eq:eq4}, the shifted energy of propagation is $p^{2}$ 
   is defined by
\begin{equation}
    p^{2}=\epsilon -E_{c} 
    \label{eq:eq5}
\end{equation}
where $\epsilon$ is the un--shifted energy 
including states below the mobility edge
$E_{c}$ which is defined as
\begin{equation}
  E_{c} = p_{c}p
    \label{eq:eq6}
\end{equation}
with
\begin{equation}
 p_{c} = 2 N_{loc}\sigma(p^{2}_{av})
    \label{eq:eq7}
\end{equation}
and $p_{av}$ is the average momentum.
   The momentum or wave number $p_{c}$ is the Kubo upper limit for diffusing
   electrons\cite{kubo}.  Finally the very important collision frequency 
   $\nu$ in eq. \ref{eq:eq4},
   which directly determines the mobility or drift velocity, is
\begin{equation}
 \nu = N \sigma v = N\sigma (p^{2}) p
    \label{eq:eq8}
\end{equation}            
where $1/N$
here is a measure of the volume over which the collision frequency
(probability) is determined, and $N_{loc} $ is the local density where 
the average momentum transfer occurs.

   It is interesting that $\hbar \nu$ is equal to the mobility edge $E_{c},$
   and also to 
   $\Gamma ,$ the imaginary part of the multiple energy shift.  
   $\Gamma$ was originally
   identified as the mobility edge in 1980 with Eq. 4 predicting the electron
   mobilities closely in He, H$_{2},$ N$_{2}$ and CO$_{2}$ with 
   no free parameters\cite{om80}.

\section{Treating the gas as an inhomogeneous medium 
(Neon)}\label{neoninhom}
   The completely unexpected observations of Borghesani {\it et 
   al.}\cite{borg90,borg88} in Neon
   have shown that the model in Sec. \ref{oldtheories} 
   (MS theory in a gas treated as
   homogeneous) is definitely overlooking something if it is to describe the
   special case of Neon as well.
   
   In fact no gas is truly a homogeneous medium.  A gas is rather a collection
   of individual atoms whose range for interacting with an electron wave is
   very small (a few atomic units).  Therefore, in the region where a
   collision actually occurs, the local average density of atoms may be more
   or less than the global average density $N.$

      There was an earlier different but equally anomalous density effect
      observed by Schwarz~\cite{schwarz80} in He.  
      As the electric field strength was
      increased, the density effect rapidly increased from very 
      negative up to zero and above.
   The effect of increasing the electric field is governed by the energy
   distribution function\cite{hux}, and the effect
   was finally understood and reproduced~\cite{om92} by recognizing first
   that the density need not be the same everywhere and second that the
   experiments were showing that the collisions  which 
   determines the energy distribution function throught the collision 
   frequency have a strong preference for occurring in regions of less 
   than average density.

  For the present case of thermal Neon mobility, Borghesani and 
  Santini\cite{borg90,borg88} noticed that adding a Fermi shift $-\Delta$
  could move the mobility strongly
  in the direction of their anomalous experimental data, and also that the
  long--wavelength limit of the structure factor $S(0)$ needed to be included.

   The theory in Sec. \ref{oldtheories}
   already incorporates the full effect of the average
   density $N .$ However, where the local density in Neon as an inhomogeneous
   medium is less than $N,$ the local density $N_{loc}$ is reduced by 
   \begin{equation}
       N_{loc}=N(1- \delta)
\label{eq:eq9}
\end{equation}
where $\delta$ is defined as $\delta = (N-N_{loc})/N.$
This further shifts the energy $p^{2}$ of Eq. \ref{eq:eq5} by the Fermi shift $\Delta$ of Eq. \ref{eq:eq2} to 
   \begin{equation}
       p^{2}=\epsilon - E_{c}+\Delta(N_{loc} -N)
\label{eq:eq10}
\end{equation}

If $N_{loc}<N,$ i.e., if $\Delta N= -N\delta$ is negative $(\delta >0),$
then $\Delta$ for Ne or He in Eq. \ref{eq:eq10} increases the energy of 
propagation $p^{2}$ and $\sigma (p^{2})$
and, by Eq. \ref{eq:eq8}, increases the collision frequency $\nu$ 
(which determines both the distribution function and the mobility). 

Thus, collisions are more probable at lower densities, and a larger 
$\nu$ also decreases the mobility $\mu,$ moving it in the direction 
of the experiment, as was noted by Borghesani and Santini\cite{borg88,borg90}. 
 (Conversely, where $\Delta N$ is positive 
the mobility would be
moved away from experiment, but this is less probable and so the smaller
densities should dominate.)

\section{Procedure}\label{proc}
   Starting from the above pressure effect model of sections 
   \ref{oldtheories} and \ref{neoninhom}, with the
   effects of multiple scattering, Kubo diffusion and the structure factor
   $S(0)$ given and the e--Ne cross sections determined\cite{OMcromp80}
   as functions of energy it remains only to explore the effect of possible
   local density deficits $\delta$ of Eq. \ref{eq:eq9} 
   on the mobilities to be calculated.
   
   In particular we look to see how the $\Delta N$'s change the mobilities and
   whether a simple choice of $\delta$'s can match all the observed mobilities as a
   function of temperature and density.  We have already
   shown how only negative values of the variation $\Delta N$ from 
   average can move
   the mobility in the direction of experiment.

   \section{Results for the mobilities in the low $E/N$ limiti}\label{results}      
 \begin{figure}[htbp]
    \centering
    \includegraphics[scale=0.5]{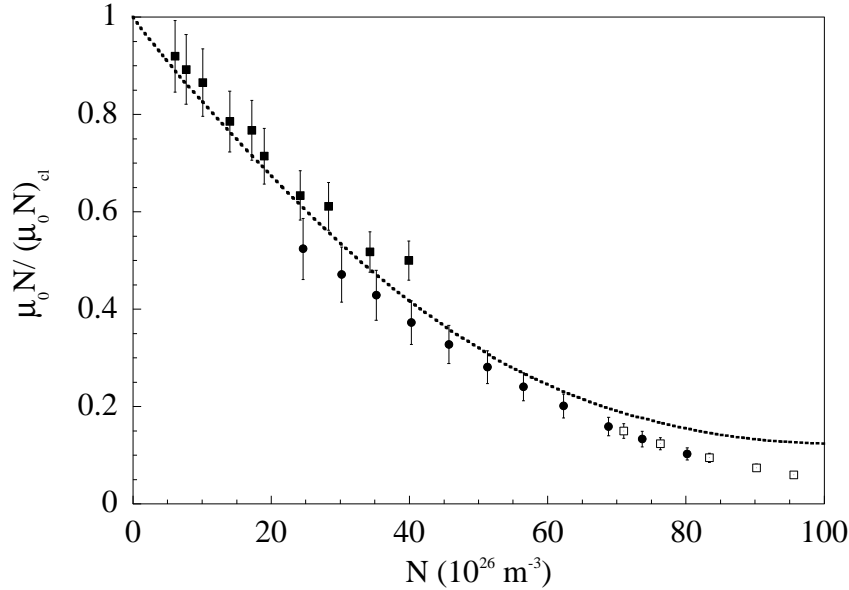}
    \caption{\small Experimental zero--field 
    density--normalized mobility ratios $(\mu_{0}N)/(\mu_{0}N)_{cl}$ data 
    in Ne gas at $T=47.9$ K (solid squares) and $T=46.5 $ K (
    solid dots and open squares)~\cite{borg88,borg90}.  Dotted line: prediction of the present one
    parameter MS theory for a inhomogeneous gas (dashed line).  The agreement
    is quite good up to a density of $\approx 70 \times 10^{26}$ 
    m$^{-3}.$}
    \label{fig:fig3}
\end{figure}
  At each temperature ($\approx 46.5-47.9,$ 77.4, 101.2, 196 and 294 K) 
  for which mobilities
  have been measured and for all moderate densities, we found a 
  single $\delta$  
  which made the predicted mobilities match the observed very closely.

   Figure \ref{fig:fig3} compares the observed pressure effect 
   $(\mu_{0}N)/(\mu_{0}N)_{cl}$ at 46.5--47.9 K to
   the present model with $\delta= 9\% .$  The fit may be seen to be
   good up to about $N=70\times 10^{26}$ m$^{-3},$ 
   where additional high density mechanisms begin to be
   important.

\begin{figure}[htbp]
    \centering
    \includegraphics[scale=0.5]{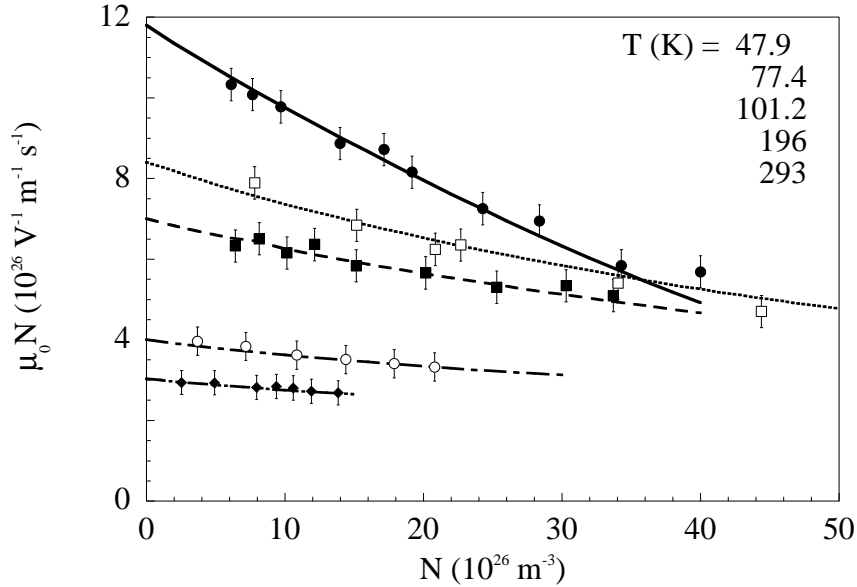}
    \caption{\small Absolute zero-field density normalized mobility 
    $\mu_{0}N$ as a
    function of the gas density in neon gas for several 
    temperatures ($T=47.9,
    77.4, 101.2, 196.0,$ and $293.0$ K, from top).  
    The lines are the predictions
    of the present one parameter MS model for an inhomogeneous gas. 
    The 101.2 K data were never published before.}
    \label{fig:fig4}
\end{figure}

      Figure \ref{fig:fig4} shows the same comparison 
      (in terms of the absolute values of $
      \mu_{0} N$at all 5
      temperatures in the moderate density range. We note that the 
      101.2 K data were never published before.

\section{Discussion}\label{dis}

\subsection{Neon vs. Helium}\label{he-ne}

Given the dramatic effect that inhomogeneities have on thermal mobilities
   in Neon, why is the same not true in He where the homogeneous gas model
   with no free parameters\cite{om80} predicts the mobilities very well?

  The answer is that the thermal mobility (Eq. \ref{eq:eq4}) 
  is determined by $1/\sigma .$  In He
  the cross section is nearly constant, while in Neon it varies rapidly with
  energy especially at the lowest energies.
  
  On the other hand, Schwarz\cite{schwarz80} 
  found a dramatic effect for the electric
  field dependence of the non--thermal mobilities which was explained by 
  a crude inhomogeneous gas model\cite{om92}.
   
   The reason that the electric field dependence in He is sensitive to density
   variations (unlike its mobility) is that, although the He cross section
   varies hardly at all, the contributions to the electron energy distribution
   are determined at small fields by $\nu^{2} = (N\sigma p)^{2}$ by Eq. 
   \ref{eq:eq4bis}.  
   So it is the
   significant shift in the momentum $p$ that causes Schwarz' effect. 
   (Unfortunately the very small size of the Neon cross sections makes this
   electric field effect barely significant in Neon).

\subsection{Temperature dependence of the density defects 
$\delta$}\label{subdis}   
\begin{figure}[htbp]
    \centering
    \includegraphics[scale=0.45]{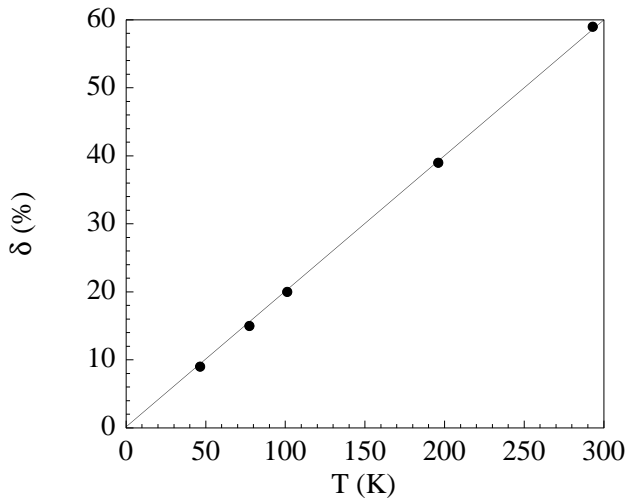}
    \caption{\small Fractional density deficit $\delta$ 
    as a function of temperature. 
    The solid line is the equation $\delta = 0.2 T.$}
    \label{fig:fig5}
\end{figure}

   Figure \ref{fig:fig5} shows the values of our empirical $\delta$ as a function of
   temperature.  They lie almost precisely on a straight line going 
   from  9\%
   at 47 K to 59\% at 294 K and vanishing at T=0.  As a result, the experiments
   at all 5 temperatures and moderate densities are all fitted closely by a
   $\delta (T)$ curve with an only single variable parameter  
\begin{equation}
\delta = 0.202\> T
\label{eq:eq11}
\end{equation}
The close agreement seems remarkable in view of the anomalous nature of the observations
and their great distance from what was previously understood. What the 
agreement seems to be saying is the following. 

According to the experiments and 
Eq. \ref{eq:eq11} the density deficit is 
proportional to $T,$ 
which is itself proportional to the inverse square of the electron's 
thermal wavelength
$\lambda_{T} = 
h/\sqrt{2m\pi k_{\mathrm{B}}T}.$ 
 It follows that $\lambda_{T}^{2} $ goes to 
infinity at $T=0 $ K and so the density sampled over such a range can only 
be the average $N.$ This means that $\delta$ must be zero  there as 
Eq. \ref{eq:eq11} predicts.
And, as $T$ increases the area $\lambda_{T}^{2}$ decreases. The 
smaller area
allows progressively larger deviations from the average density $N$ 
also as Eq. \ref{eq:eq11} would predict.

But what is the significance of the area $\lambda_{T}^{2}$? The interaction of an electron 
with an atom is known to be limited to a distance no greater than a 
wavelength. Thus, as the electron's wavepacket advances forward, $\lambda_{T}^{2}$
may be interpreted as a measure of the largest area 
 over which it can overlap 
with the e--atom interaction and sample the local density.

\subsection{The quantum inferences}\label{qc}

   The theory used in this work starts with pure Boltzmann theory at very low
   density, but uses quantum cross sections for the important
   momentum-transfer collision frequency~\cite{hux}.  It	
   describes an electron moving classically in a well defined direction from
   one momentum/energy transfer to the next.

   However, in light of quantum theory, we know that the electronÕs motion is
   physically described by a wave function spreading out broadly, with Feynman
   paths passing through every atom in its forward direction.  The total wave
   may alternately be represented as a superposition of wave packets, each
   beginning at the electron's starting point and ending on a possible 
   interaction with an individual atom
   in its path.  For each wave packet, there is a well defined momentum
   vector in between the starting point and any atom.

   What the earlier close agreement of the classical Boltzmann theory with the
   low density experiments indicates is that, after a momentum/energy transfer
   with one atom, the electron moves forward starting from that one point. In
   other words all the remaining members of the original superposition become
   irrelevant as the electron must be continuing on from the point of momentum transfer
   as a new wave.
   
   The narrowing of the original superposition plus the electron 
   continuing from that point closely parallels Keller's 
   derivation\cite{keller} of wave function collapse in which he 
   proved, using conditional probability theory, that a direct 
   observation results in a collapsed single component wave--function 
   at the observation point ready to continue from there.
   
   What these results add to Keller's conclusion is that observation 
   may be generalized to momentum transfer interaction. More detailed 
   evidence of this collapse to something like a particle has been 
   seen at higher densities, particularly in neon gas.

   When the density $N$ was increased in electron mobility experiments, known
   density dependent mechanisms were recognized as influencing the outcome,
   the principal one being the multiple scattering of waves.

   What this mechanism does first is to modify the electron wave--function's
   kinetic energy, and therefore its collision frequency $\nu$ 
   (Eq.\ref{eq:eq8}) by the
   density dependent kinetic energy shift $ \Delta$ (Eq.  \ref{eq:eq2}).  
   ($-\Delta$ is the cumulative
   potential energy of the electron summed over all the atoms in the full wave
   propagating wave-function).
    
   This potential scattering, resulting from multiple scattering, is in
   addition to the rare energy exchange at the actual momentum transfers.  The
   latter is completely determined by $\nu ,$ as well as the energy balance,
   through $g(\epsilon ),$ (Eq.  5).
   
\subsection{The present Neon case}\label{prene}

With Neon as the gas, the fact that its scattering cross section is close
to zero and changing rapidly makes the collision frequency exceptionally
sensitive to gas density where momentum transfers occur. This 
sensitivity enabled us, as discussed above, to explain and predict the 
anomalous density effect in Neon and also to discover that these 
transfers were occurring mostly at local densities less than the 
average one $N$ by an amount proportional to the temperature.

 The fact that the present model's densities at points of momentum
   transfer were found to be significantly different from the average 
   $N$ implies that the individual wave packet at the interaction, even before moving
   forward, is more compact than the full wave packet during its
   propagation - making it look much more like a ÒparticleÓ there.

   This presently inferred shrinking or collapse of the electron's 
   extended wave packet at the point of momentum transfer is further 
   evidence of the behavior of the wave function demonstrated by 
   Keller about the point of observation.
      
Our finding from the experiments that the density defect $\delta$ is 
such that at 
most momentum transfers $N_{loc}$ can differ from the average $N$ by 
as much as 59 $\%$ accordingly shows that the wave function for an 
electron's motion, where it interacts, is very much smaller than the 
free propagating wave function. It might understandably be called a 
{\it particle} -- as in the photoelectric effect, where light waves 
become point--like photons when they exchange energy and momentum with 
electrons in a metal, or the way photons and electrons were described 
earlier when they were emitted or observed.

   Where the present finding from Neon mobilities differs from 
   Keller's is first that it is the wave and wave packet for electron 
   motion which shrinks and second that the collapse--like behavior 
   happens at every momentum/energy transfer whether or not it is 
   observed directly.

\section{Summary}\label{sec:concl}

A straight-forward extension of existing theoretical models, combining the
inhomogeneous nature of a gas with multiple scattering theory, has been
found to reproduce the anomalous and previously puzzling measurements of
the pressure effect on electron mobilities in Neon gas at 5 temperatures
and moderate densities both simply and accurately with only one variable
parameter - the deviation $\delta$ (proportional to the temperature) of local
from average gas density at actual momentum transfers.

   The special window that these, and also previous, electron mobility
   experiments offer on fundamental quantum processes was also discussed.

   Further experimental investigations of this kind of relation between
   collisions in a gas and electron (or atom wave) packets are strongly
   recommended.

\section*{Acknowledgments}
Dr. Howard Hanley provided valuable assistance with the Neon structure 
factor and the support of Contractors Register is gratefully 
acknowledged. A.F.B. would like to thank Prof. M.Santini for 
invaluable discussions. T.O'M. would like to dedicate his contribution to his 
late wife Tanya. 
%


\end{document}